\documentclass[conference]{IEEEtran}
\IEEEoverridecommandlockouts

\usepackage{cite}
\usepackage{amsmath,amssymb,amsfonts}
\usepackage{algorithmic}
\usepackage{graphicx}
\usepackage{textcomp}

\usepackage{booktabs}       
\usepackage{multirow}
\usepackage{float}
\usepackage[dvipsnames]{xcolor}
\usepackage{colortbl}
\usepackage{bm}
\usepackage{arydshln}
\usepackage{makecell}
\usepackage{siunitx}

\def\BibTeX{{\rm B\kern-.05em{\sc i\kern-.025em b}\kern-.08em
    T\kern-.1667em\lower.7ex\hbox{E}\kern-.125emX}}

\begin{document}

\title{Improving Resource-Efficient Speech Enhancement via Neural Differentiable DSP Vocoder Refinement\\
\thanks{$\dagger$~Work done during an internship at Meta (Reality Labs Research).\\ Corresponding author e-mail: heitor.guimaraes@inrs.ca}
}

\makeatletter
\newcommand{\linebreakand}{%
  \end{@IEEEauthorhalign}
  \hfill\mbox{}\par
  \mbox{}\hfill\begin{@IEEEauthorhalign}
}
\makeatother

\author{\IEEEauthorblockN{Heitor R. Guimarães~$^\dagger$}
\IEEEauthorblockA{\textit{INRS - EMT} \\
Montreal, CA}
\and
\IEEEauthorblockN{Ke Tan}
\IEEEauthorblockA{\textit{Meta Reality Labs} \\
Redmond, US}
\and
\IEEEauthorblockN{Juan Azcarreta}
\IEEEauthorblockA{\textit{Meta Reality Labs} \\
Cambridge, UK}
\and
\IEEEauthorblockN{Jesus Alvarez}
\IEEEauthorblockA{\textit{Meta Reality Labs} \\
Castile Y Leon, ES}
\linebreakand
\IEEEauthorblockN{Prabhav Agrawal}
\IEEEauthorblockA{\textit{Meta AI} \\
New York, US}
\and
\IEEEauthorblockN{Ashutosh Pandey}
\IEEEauthorblockA{\textit{Meta Reality Labs} \\
Redmond, US}
\and
\IEEEauthorblockN{Buye Xu}
\IEEEauthorblockA{\textit{Meta Reality Labs} \\
Redmond, US}
}

\maketitle

\begin{abstract}
Deploying speech enhancement (SE) systems in wearable devices, such as smart glasses, is challenging due to the limited computational resources on the device. Although deep learning methods have achieved high-quality results, their computational cost limits their feasibility on embedded platforms. This work presents an efficient end-to-end SE framework that leverages a Differentiable Digital Signal Processing (DDSP) vocoder for high-quality speech synthesis. First, a compact neural network predicts enhanced acoustic features from noisy speech: spectral envelope, fundamental frequency ($\mathbf{F}_0$), and periodicity. These features are fed into the DDSP vocoder to synthesize the enhanced waveform. The system is trained end-to-end with STFT and adversarial losses, enabling direct optimization at the feature and waveform levels. Experimental results show that our method improves intelligibility and quality by 4\% (STOI) and 19\% (DNSMOS) over strong baselines without significantly increasing computation, making it well-suited for real-time applications.
\end{abstract}


\section{Introduction}
Real-world acoustic environments often contain noise and reverberation, which significantly degrade speech intelligibility and hinder effective communication. Speech enhancement (SE) systems aim to suppress such detrimental factors and isolate the target speech signal, boosting the perceived quality and/or intelligibility~\cite{loizou2007speech}. The emergence of wearable devices, such as smart glasses, brings SE technologies closer to everyday use. However, practical deployment remains challenging due to hardware constraints of these platforms, including limited computational resources, strict power budgets, and latency requirements~\cite{pandey2023simple, 10446087}. While deep learning-based SE methods have achieved state-of-the-art performance in controlled settings~\cite{10890512, cheng2024dynamic, 10889061}, ensuring robust operation on resource-limited devices in unconstrained environments remains an open challenge.


Deep learning-based SE methods are typically categorized into generative and discriminative approaches, with the latter recently achieving state-of-the-art performance for perceptual quality and fidelity~\cite{xue2024low, yang24h_interspeech, guimaraes2025ditse, scheibler24_interspeech, lee2025flowse}. However, generative models are often computationally intensive and parameter-heavy, limiting their practicality for real-time or low-resource scenarios. Discriminative-based SE methods are commonly framed as regression problems, where models are optimized to minimize an $L_p$ norm between the enhanced and target clean speech. However, these losses act as surrogate objectives and are poorly correlated with perceptual quality or intelligibility as experienced by human listeners. Consequently, while these methods may achieve low reconstruction error, they often fail to produce speech that sounds natural or is easy to understand, particularly in adverse acoustic environments.


To bridge this perceptual gap, recently neural vocoders for have been used as a post-processing tool to boost quality. Models such as HiFi-GAN~\cite{kong2020hifi} and Vocos~\cite{siuzdak2024vocos} have been increasingly used to improve perceptual quality by synthesizing natural-sounding speech from intermediate representations (e.g., spectrograms or self-supervised learned features~\cite{10715278}). A typical two-stage pipeline first denoises the spectrogram, and then does the waveform reconstruction. Using neural vocoders often leads to high computational cost, limiting their use in resource-constrained settings like wearables. Differentiable Digital Signal Processing (DDSP)~\cite{Engel2020DDSP}, on the other hand, integrates strong inductive biases from speech production, signal processing, and perception into deep learning models with a low computational overhead, making it ideal for resource-constrained scenarios. Human speech is generated from harmonic (voiced) and noise-like (unvoiced) excitations\cite{o1987speech}, with perceptual quality and intelligibility relying on accurate dynamic representation of harmonics and spectral transitions near consonant boundaries~\cite{10382416}. Traditional SE methods often overlook these phonetic aspects. In contrast, the DDSP vocoder explicitly models these components, leading to improved generalization and perceptual quality.

\begin{figure*}[t]
        \centering
        \includegraphics[width=\linewidth]{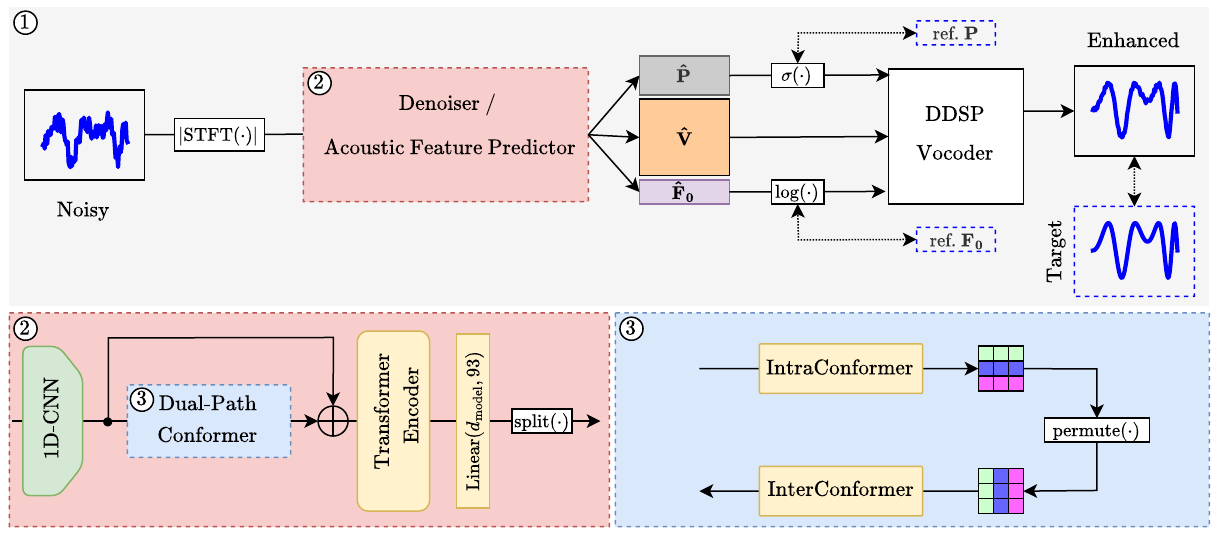}
        \caption{Overview of the proposed system. The acoustic feature predictor estimates periodicity, spectral envelope, and fundamental frequency from noisy input, which the DDSP vocoder uses to synthesize enhanced speech. In (1), white boxes denote non-parametric operations; dashed blue boxes indicate training-only supervision; dotted arrows mark loss application points. (2) and (3) provide detailed views of the feature predictor and dual-path conformer, respectively.}
        \label{fig:ddsp_se_arch}
\end{figure*}

In this work, we explore the use of the DDSP vocoder~\cite{10447948} as a core component for speech enhancement. Our method follows a two-stage architecture: a lightweight neural network first estimates enhanced acoustic features from noisy input audio; these features are then passed to a DDSP vocoder to synthesize the enhanced waveform. Leveraging the fully differentiable nature of the DDSP architecture, our system is trained in an end-to-end manner using a combination of adversarial and reconstruction losses, enabling optimization directly on the output waveform and at the acoustic feature level. The closest related work is the Neural Homomorphic Vocoder (NHV) series~\cite{liu20_interspeech, jiang22b_interspeech, 9747177, 10109754}, which employs a DSP-based vocoder for enhancement synthesis. Compared to NHV, the DDSP vocoder achieves superior audio quality with a 24$\times$ reduction in computation by using a zero-phase DDSP filter design. This eliminates the need for explicit phase modeling, simplifies synthesis, and reduces artifacts through separate modeling of periodic and aperiodic components. In the enhancement application, despite being three times smaller than the most compact NHV variant (roughly 1M parameters), our model delivers competitive results, making it well-suited for resource-constrained applications.

\section{Proposed Method}

This section outlines the proposed methodology for extracting enhanced features from a noisy input utterance with a low-complexity neural network and resynthesizing the enhanced speech with the DDSP vocoder, as illustrated in Figure~\ref{fig:ddsp_se_arch}.

\subsection{Problem formulation}

Consider a single-channel microphone capturing an audio mixture $\mathbf{y} \in \mathbb{R}^{N}$, where $N$ is the number of samples. This signal can be decomposed into a clean speech signal $\mathbf{s} \in \mathbb{R}^{N}$, background noise $\mathbf{n} \in \mathbb{R}^{N}$, and a room impulse response $\mathbf{h} \in \mathbb{R}^{N}$ that inserts the room reverberation in the mixture. Mathematically, this mixture can be expressed as $\mathbf{y} = \mathbf{s} \ast \mathbf{h} + \mathbf{n}$, where $\ast$ denotes the linear convolution operator.

\begin{figure}[htpb]
        \centering
        \includegraphics[width=\linewidth]{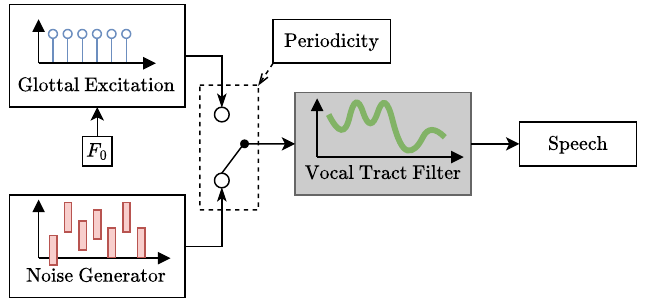}
        \caption{Simplified schema for the Source-Filter~\cite{fant1971acoustic} model.}
        \label{fig:source_filter}
\end{figure}

Speech enhancement aims to minimize a distance metric between the enhanced signal $\hat{\mathbf{s}}$ and a reference signal $\mathbf{s}$.
As previously discussed, vocoders offer a promising approach to incorporating speech-production inductive bias into our enhancement model, thereby generating high-quality outputs. In particular, source-filter-based vocoder methods synthesize speech from acoustic parameters as shown in Figure~\ref{fig:source_filter}. Our goal is to accurately predict clean acoustic parameters from noisy speech using a neural network, and then resynthesize the enhanced speech through the vocoder, as detailed in the following subsection. 

\subsection{Speech Synthesis via DDSP Vocoder}

Human speech is generated through two main excitation mechanisms: quasi-periodic vocal fold vibrations, modulated by the fundamental frequency ($F_0$), producing voiced sonorants with harmonics shaped by vocal tract (VT) resonances; and noise-like excitations from turbulent airflow at narrow constrictions, which create obstruents~\cite{o1987speech}. Furthermore, perceptual quality and intelligibility largely depend on two acoustic cues: accurate dynamic representation of harmonics and spectral transitions near consonant boundaries~\cite{10382416}.

In the source-filter model for speech production, the excitation signal $\mathbf{E}$ and the vocal tract filter $\mathbf{V}$ are the main components to synthesize an utterance. The DDSP Vocoder~\cite{10447948} proposes to further decompose the excitation signal into a periodic (impulse train) and an aperiodic (white noise) signal of same energy by multiplying them with the periodicity feature. Furthermore, three features are necessary for its implementation: (i) Fundamental frequency ($\mathbf{F_0}$), which is a 1-dimensional vector; (ii) Periodicity ($\mathbf{P}$), a $p$-dimensional mel-bandwise ratio between periodic and aperiodic excitation; and (iii) the vocal tract filter ($\mathbf{V}$), a $v$-dimensional linear frequency magnitude tensor. The synthesized speech $\mathbf{s}$ can be written as

\begin{equation}
    \mathbf{s} = \underbrace{\text{iFFT}(\mathbf{P} \times \mathbf{V}) \ast e_{\text{imp}}(\mathbf{F_0})}_{\text{Periodic signal}}  +  \underbrace{\text{iFFT}((1-\mathbf{P}) \times \mathbf{V} \times E_{\text{noise}})}_{\text{Aperiodic signal}},
\end{equation}
where $e_{\text{imp}}$ is the glottal excitation, an impulse train in the time-domain modulated by fundamental frequency; and $E_{\text{noise}}$ is a white noise signal in the frequency-domain. 

In summary, we adopt the DDSP vocoder for its end-to-end differentiability and efficiency due to the lack of trainable parameters (i.e., no neural network in the vocoder itself).

\subsection{Input features}

The input signal $\mathbf{y}$ is transformed into the time-frequency domain via the Short-Time Fourier Transform (STFT), resulting in a complex-valued tensor $\mathbf{Y} \in \mathbb{C}^{T \times F}$, where $T$ is the number of frames, and $F$ is the number of frequency bins. We compute the magnitude spectrogram $|Y|$ and apply a mel-filterbank transformation to project it onto a lower-dimensional, perceptually motivated frequency scale. This operation reduces the frequency resolution by aggregating adjacent bins, particularly in the high-frequency range.

\subsection{Locally-connected low-level features}

The first processing step in our network is to be able to address the local structure and dependencies of speech signals. Therefore, we employ a small CNN with a similar structure to the Descript Audio Codec (DAC) encoder~\cite{kumar2023high}. First, a weight-normalized (WN) 1D-convolution layer is applied, with four channels, a kernel size of three, and same padding. Each frequency bin of the mel-spectrogram is treated as an individual channel in this step. Next, three units containing dilated convolutions with dilation rates of (1, 3, 5) are applied. Each unit consists of a Snake1D~\cite{ziyin2020neural, lee2023bigvgan} activation function, a dilated WN-1D-convolution, an activation function again, and a final WN-1D-convolution with kernel size 1. The number of channels and stride is constant and equal to 3 and 1, respectively. A final Snake1D activation and a WN-1D-convolution layer are applied in this step. Lastly, a ReLU activation function and a LayerNorm operation are applied before the input to the Transformer-based network.

\subsection{Time-Frequency processing with the Dual-Path Conformer}

The last components of our approach, before the DDSP vocoder, is the temporal processing module. Inspired by~\cite{subakan2021attention}, we leverage a dual-path conformer network for the denoising of the received feature map, looking to both the temporal and frequency axes. The dual-path conformer estimates a gain that is applied in the original feature map, hence denoising in the latent space. After this, we use a small transformer network with $L$ layers to estimate the acoustic features.

Its output is passed through a linear projection layer that maps the hidden dimension $d_{\text{model}}$ to $v + p + 1$, where $v$ corresponds to the number of frequency bins in the spectral envelope, $p$ to the number of periodicity bands, and the scalar $1$ represents the predicted fundamental frequency ($\mathbf{\hat{F}}_0$). This projected output is then split into three separate acoustic features: spectral envelope ($\mathbf{\hat{V}}$), periodicity ($\mathbf{\hat{P}}$), and ($\mathbf{\hat{F}_0}$), which are used as inputs to the DDSP vocoder. Following the original DDSP setup, we set $v=80$ and $p=12$. The spectral envelope is further converted from log Mel-scale (80-dim) to a 129-dim linear-frequency magnitude representation for synthesis.

\subsection{Loss function}

Three components are optimized in our model: The predicted $\mathbf{\hat{F}}_0$, periodicity $\mathbf{\hat{P}}$, and the final synthesized speech signal itself. For both $F_0$ and periodicity, a mean squared error (MSE) between the predicted and the reference extracted from the clean signal is computed. For the re-synthesized speech signal, we compute the multi-resolution STFT loss~\cite{yamamoto2020parallel, defossez2020real}, with window lengths of $W = \{512, 1024, 2048\}$ samples, a fixed stride of 8 ms, and loss weights of $\Lambda = \{25.7, 51.3, 102.5\}$, respectively. Furthermore, the last component is an adversarial loss, which follows the original DDSP vocoder implementation, leveraging $K=16$ discriminators operating on the 129-dim magnitude spectrograms, extracted at \SI{8}{ms} hop size, with a least square adversarial loss formulation.

Finally, the generator (enhancement system) loss function is computed as the weighted sum of all components as

\begin{equation}
    \mathcal{L}^{(G)} = \alpha\mathcal{L}_2(\mathbf{\hat{F}}_0, \mathbf{F}_0) + \beta\mathcal{L}_2(\mathbf{\hat{P}}, \mathbf{P}) + \mathcal{L}_\text{MRS}(\mathbf{\hat{s}}, \mathbf{s}) + \mathcal{L}^{(G)}_\text{adv}(\mathbf{\hat{s}}),
\end{equation}
where $\mathcal{L}_2(\mathbf{\hat{x}}, \mathbf{x}) = ||\mathbf{\hat{x}} - \mathbf{x}||_2$ is the MSE loss function, $\alpha$ and $\beta$ are hyperparameters, and $\mathcal{L}_\text{MRS}$ is the multi-resolution STFT loss defined as

\begin{equation}
    \mathcal{L}_\text{MRS}(\mathbf{\hat{s}}, \mathbf{s}) = \sum_{\substack{w_i \in W\\ \lambda_i \in \Lambda}}  \lambda_i\frac{||f(\mathbf{\hat{s}}, w_i) - f(\mathbf{s}, w_i)||_1}{||f(\mathbf{\hat{s}}, w_i)||_1},
\end{equation}
with $f(\mathbf{x}, w_i) = \log{|\text{STFT}^{(w_i)}(\mathbf{x})|}$ being the log magnitude-spectrogram of the signal $\mathbf{x}$ with a window size $w_i$ and a fixed hop length 8ms. Lastly, the adversarial loss for the generator, $\mathcal{L}^{(G)}_\text{adv}(\mathbf{\hat{s}}, \mathbf{s})$, can be written as
\begin{equation}
    \mathcal{L}^{(G)}_\text{adv}(\mathbf{\hat{s}})= \sum_{k=1}^{K}\frac{D_k(\mathbf{\hat{s}})-1}{N_k},
\end{equation}
where $D_k$ is the $k^{\text{th}}$ sub-band discriminator operating over a specific STFT magnitude band, and $ N_k$ represents the corresponding number of spectral elements within that band.

\section{Experimental Setup}

\subsection{Dataset}
In this work, we leverage the Interspeech 2020 DNS Challenge corpus~\cite{reddy20_interspeech} to create degraded and clean audio pairs resampled to 16 kHz. Room impulse responses (RIRs) are synthesized using the Pyroomacoustics library, which models acoustic propagation within a three-dimensional shoebox-shaped environment. The room dimensions are randomized for each simulation: the length and width are independently drawn from a uniform distribution over the interval [3, 10] meters, while the height is sampled from [3, 4] meters. To introduce variability in reverberation characteristics, the surface absorption coefficient is also uniformly sampled within the range [0.6, 0.95]. This procedure yields diverse reverberant conditions for robust evaluation of model performance under real-world acoustic scenarios.

Herein, the microphone position and orientation are randomly assigned within the room. Between 1 and 10 noise sources are positioned at distances greater than 0.5 meters. To simulate babble noise, 5–20 interfering talkers are introduced with 75\% probability, all located beyond 3 meters. Signal-to-Noise Ratio (SNR) and Signal-to-Interference Ratio (SIR) are uniformly sampled from [–5, 10]~dB and [5, 10]~dB, respectively.

In summary, we use a subset of the English dataset, randomly selecting 128K samples for training, 1000 for validation, and 500 for testing. Each train and validation sample has a duration of \SI{10}{s}, while the test samples have \SI{20}{s}.

\subsection{Model architecture and Baselines}

In this work, we explore two baseline architectures upon which we integrate our post-processing DDSP vocoder. First, we consider a Convolutional Transformer Network (CTN), a non-causal model composed of three main stages: a convolutional front-end, a dual-path Conformer module, and a Transformer encoder, as described in detail in the previous section. We consider two variants of the CTN architecture: a compact model with approximately 300K parameters and a larger counterpart with around 600K parameters. In the smaller configuration, the dual-path network comprises an intra- and inter-conformer module with a single layer and a model dimension $d_{\text{model}}=64$, eight attention heads, a feedforward network of dimension 96, and a convolutional kernel size of 7. In the larger configuration, the intra- and inter-conformer modules maintain the same $d_{\text{model}}$, number of attention heads, and kernel size, while increasing the number of layers to four, and the feedforward dimension to 128.

Following the dual-path Conformer, both model variants incorporate a Transformer encoder consisting of 4 layers, each with 8 attention heads, a feedforward network of dimension 96, and the same $d_{\text{model}}=64$ as in the preceding Conformer modules. The output of the Transformer is passed through a ReLU activation, followed by a linear projection layer that maps the feature dimension from $d_{\text{model}}$ to 93. This output dimensionality corresponds to the predicted acoustic parameters the DDSP vocoder requires: 80 dimensions for the spectral envelope, 12 dimensions for the periodicity, and 1 dimension for the fundamental frequency.

In addition to the CTN architecture, we investigate a Convolutional Recurrent Network (CRN) following the design proposed in~\cite{8492428}. The CRN is a compact, causal U-Net-style architecture composed of an encoder and decoder implemented with 2D convolutional and transposed convolutional layers, respectively, and a recurrent bottleneck module based on Gated Recurrent Units (GRUs).

As with the CTN, we evaluate two model configurations: a compact variant with approximately 300K parameters and a larger variant with roughly 600K parameters. Both configurations employ 4-layer encoder and decoder blocks. In the compact configuration, the encoder consists of convolutional layers with channel dimensions $[1, 8, 16, 32]$, using a kernel size of $(2, 3)$, and stride of $(1, 2)$. The bottleneck comprises a two-layer GRU with a hidden size of 24, while the decoder consists of transposed convolutional layers with channel dimensions $[24, 32, 16, 8]$. In the large configuration, the primary differences lie in the bottleneck and decoder. The GRU bottleneck is expanded to three layers with a hidden size of 32, and the decoder is modified to use channel dimensions $[32, 32, 16, 8]$, providing increased model capacity.

When integrating the DDSP vocoder with the CRN backbone, we append a lightweight Transformer network composed of 4 layers, with $d_{\text{model}}=64$ and a feedforward dimension of 96. Unlike the Transformer used in the CTN architecture, this variant is strictly causal; therefore, each attention mechanism is constrained to attend only to past and present positions to preserve the overall causality of the system.

All models are trained using a batch size of 64 utterances and the Adam~\cite{kingma2014adam} optimizer. For models incorporating the DDSP vocoder, the learning rate is initially set to $10^{-4}$ and linearly warmed up to $10^{-3}$ over the first five epochs, which precedes the beginning of the discriminator training. Following this, the learning rate decays linearly to $10^{-5}$ over the remaining 245 epochs, resulting in a total of 250 training epochs. In contrast, models that do not employ the DDSP vocoder are trained for 100 epochs using a constant learning rate of $10^{-4}$, which was found to be sufficient for convergence and yielded the best performance for baseline configurations in our ablations.

\begin{table*}[!htb]
    \centering
    \caption{Mean and standard deviation (across samples) results of the Transformer-based model with and without the DDSP vocoder, when the input is a magnitude- or complex-STFT representation. \textbf{Bold} numbers highlight the best achieved metric.}
    \label{tab:siso_experiments}
    \resizebox{\linewidth}{!}{%
        \begin{tabular}{lccccccccc}
            \toprule
            \multirow{2}{*}{Method} & \multirow{2}{*}{\#P (K)} & \multirow{2}{*}{PESQ ($\uparrow$)} & \multirow{2}{*}{STOI ($\uparrow$)} & \multicolumn{3}{c}{DNSMOS ($\uparrow$)} & \multirow{2}{*}{MCD ($\downarrow$)} & \multirow{2}{*}{SSIM ($\uparrow$)} \\
            \cmidrule{5-7}
            & & & & SIG & BAK & OVRL & \\

            \midrule
            Input & --- & 1.46 {\scriptsize \textcolor{gray}{$\pm$ 0.32}} & 65.78 {\scriptsize \textcolor{gray}{$\pm$ 8.03}} & 1.36 {\scriptsize \textcolor{gray}{$\pm$ 0.40}} & 1.17 {\scriptsize \textcolor{gray}{$\pm$ 0.10}}  & 2.35 {\scriptsize \textcolor{gray}{$\pm$ 0.19}} & 20.64 {\scriptsize \textcolor{gray}{$\pm$ 3.71}} & 0.46 {\scriptsize \textcolor{gray}{$\pm$ 0.15}} \\

            \midrule
            \multicolumn{9}{c}{Magnitude-STFT Input} \\
            \midrule

            \multirow{2}{*}{CTN} & 291 & 1.93 {\scriptsize \textcolor{gray}{$\pm$ 0.29}} & 73.00 {\scriptsize \textcolor{gray}{$\pm$ 8.44}} & 2.72 {\scriptsize \textcolor{gray}{$\pm$ 0.30}} & 3.05 {\scriptsize \textcolor{gray}{$\pm$ 0.34}} & 2.76 {\scriptsize \textcolor{gray}{$\pm$ 0.23}} & 6.85 {\scriptsize \textcolor{gray}{$\pm$ 1.31}} & 0.81 {\scriptsize \textcolor{gray}{$\pm$ 0.05}} \\
            
            & 586 & 2.02 {\scriptsize \textcolor{gray}{$\pm$ 0.27}} & 75.10 {\scriptsize \textcolor{gray}{$\pm$ 7.94}} & 2.81 {\scriptsize \textcolor{gray}{$\pm$ 0.29}} & 3.17 {\scriptsize \textcolor{gray}{$\pm$ 0.33}} & 2.83 {\scriptsize \textcolor{gray}{$\pm$ 0.24}} & 6.43 {\scriptsize \textcolor{gray}{$\pm$ 1.24}} & 0.82 {\scriptsize \textcolor{gray}{$\pm$ 0.05}}  \\

            \hdashline

            \multirow{2}{*}{(\textbf{Ours}) CTN + DDSP} & 307 & 2.03 {\scriptsize \textcolor{gray}{$\pm$ 0.27}} & 75.29 {\scriptsize \textcolor{gray}{$\pm$ 7.34}} & 3.23 {\scriptsize \textcolor{gray}{$\pm$ 0.24}} & 3.79 {\scriptsize \textcolor{gray}{$\pm$ 0.26}} & 3.15 {\scriptsize \textcolor{gray}{$\pm$ 0.22}}  & 5.89 {\scriptsize \textcolor{gray}{$\pm$ 1.07}} & 0.83 {\scriptsize \textcolor{gray}{$\pm$ 0.04}}\\
            
            & 594 & \textbf{2.14} {\scriptsize \textcolor{gray}{$\pm$ 0.28}} & \textbf{78.25} {\scriptsize \textcolor{gray}{$\pm$ 7.01}} & \textbf{3.43} {\scriptsize \textcolor{gray}{$\pm$ 0.17}} & \textbf{3.92} {\scriptsize \textcolor{gray}{$\pm$ 0.20}} & \textbf{3.37} {\scriptsize \textcolor{gray}{$\pm$ 0.24}}  & \textbf{5.50} {\scriptsize \textcolor{gray}{$\pm$ 1.00}} & \textbf{0.84} {\scriptsize \textcolor{gray}{$\pm$ 0.04}} \\

            \midrule
            \multicolumn{9}{c}{Complex-STFT Input} \\
            \midrule

            \multirow{2}{*}{CTN} & 299 & 2.01 {\scriptsize \textcolor{gray}{$\pm$ 0.28}} & 74.94 {\scriptsize \textcolor{gray}{$\pm$ 8.14}} & 2.69 {\scriptsize \textcolor{gray}{$\pm$ 0.30}} & 3.11 {\scriptsize \textcolor{gray}{$\pm$ 0.33}} & 2.81 {\scriptsize \textcolor{gray}{$\pm$ 0.23}} & 6.49 {\scriptsize \textcolor{gray}{$\pm$ 1.30}} & 0.82 {\scriptsize \textcolor{gray}{$\pm$ 0.05}} \\

            & 602 & 2.02 {\scriptsize \textcolor{gray}{$\pm$ 0.29}} & 75.41 {\scriptsize \textcolor{gray}{$\pm$ 8.12}} & 2.69 {\scriptsize \textcolor{gray}{$\pm$ 0.31}} & 3.28 {\scriptsize \textcolor{gray}{$\pm$ 0.33}} & 2.83 {\scriptsize \textcolor{gray}{$\pm$ 0.24}} & 6.20 {\scriptsize \textcolor{gray}{$\pm$ 1.18}} & 0.83 {\scriptsize \textcolor{gray}{$\pm$ 0.05}} \\

            \hdashline

            \multirow{2}{*}{(\textbf{Ours}) CTN + DDSP} & 307 & 2.02 {\scriptsize \textcolor{gray}{$\pm$ 0.27}} & 74.93 {\scriptsize \textcolor{gray}{$\pm$ 7.50}} & 3.20 {\scriptsize \textcolor{gray}{$\pm$ 0.26}} & 3.83 {\scriptsize \textcolor{gray}{$\pm$ 0.23}} & 3.11 {\scriptsize \textcolor{gray}{$\pm$ 0.22}} & 5.84 {\scriptsize \textcolor{gray}{$\pm$ 1.04}} & 0.83 {\scriptsize \textcolor{gray}{$\pm$ 0.04}}  \\
            & 594 & 2.05 {\scriptsize \textcolor{gray}{$\pm$ 0.26}} & 76.26 {\scriptsize \textcolor{gray}{$\pm$ 6.89}} & 3.30 {\scriptsize \textcolor{gray}{$\pm$ 0.21}} & 3.86 {\scriptsize \textcolor{gray}{$\pm$ 0.22}} & 3.22 {\scriptsize \textcolor{gray}{$\pm$ 0.22}} & 5.54 {\scriptsize \textcolor{gray}{$\pm$ 0.98}} & 0.83 {\scriptsize \textcolor{gray}{$\pm$ 0.04}} \\

            \midrule[\heavyrulewidth]
            \bottomrule
        \end{tabular}
    }
\end{table*}

\subsection{Evaluation Metrics}\label{sec:fig-of-merit}

To comprehensively assess the performance of our methodology, we employ intrusive and non-intrusive metrics that capture speech quality and intelligibility. In addition, we also incorporate two standard evaluation metrics from speech synthesis literature~\cite{9746698} to better quantify perceptual fidelity.

\begin{enumerate}
    \item \textbf{PESQ}~\cite{rix2001perceptual} and \textbf{STOI}~\cite{taal2011algorithm}: We leverage classical intrusive metrics for speech enhancement, namely Perceptual Evaluation of Speech Quality (PESQ) and Short-Time Objective Intelligibility (STOI). PESQ estimates perceived speech quality by comparing the enhanced signal with a clean reference, with scores ranging from -0.5 to 4.5. In particular, in this work, we use the narrowband version of the metric. STOI assesses speech intelligibility by measuring short-time spectral correlations, producing normalized scores between 0 and 100. For both metrics, higher values indicate better performance.

    \item \textbf{DNSMOS}~\cite{reddy2021dnsmos}: We evaluate perceptual speech quality using the DNSMOS, a non-intrusive, deep learning-based metric that estimates mean opinion scores (MOS). In particular, we use the P.808 variant, which is trained to align with ITU-T P.808 subjective evaluation standards and provides separate scores for speech quality (SIG), background noise quality (BAK), and overall quality (OVL). Higher scores indicate better perceived quality.

    \item \textbf{MCD}~\cite{9383567}: The mel cepstral distortion (MCD) computes the difference between the mel cepstral coefficients of the enhanced and reference signals. We use a slightly different methodology that first applies a dynamic time warping (DTW) algorithm to ensure alignment between the enhanced and reference signals. Smaller values indicate smaller distortion and, thus, higher similarity.

    \item \textbf{SSIM}~\cite{wang2004image, 9746698}:  The Structural Similarity Index Measure (SSIM) was initially developed to assess the similarity between two images quantitatively. Similarly to other works in the speech synthesis domain,  we convert the signals into spectrograms and treat them as images to be able to apply the metric. In this metric, higher values indicate higher similarity to the clean speech. 
\end{enumerate}

\section{Results}

\subsection{Analysis of the CTN model for a non-causal scenario}

We begin by evaluating the impact of incorporating the DDSP vocoder into the CTN model under a non-causal setting. Table~\ref{tab:siso_experiments} presents results for two input configurations: (i) magnitude-only STFT inputs, which are subsequently transformed into Mel-spectrograms, and (ii) complex-valued STFT inputs, which retain phase information. This comparison allows us to assess whether access to phase contributes to improved generalization and overall enhancement quality.

For models using real-valued inputs, we observe that increasing model size consistently improves performance, as expected in Transformer-based architectures. Incorporating the DDSP vocoder yields slight gains in the intrusive PESQ and STOI metrics. However, substantial improvements are observed in DNSMOS, a non-intrusive perceptual quality metric. We hypothesize that intrusive metrics like PESQ and STOI may not reliably capture the perceptual quality of generative models, as small misalignments between the generated and reference signals can negatively affect scores. Additionally, minor vocoder-induced artifacts may impact these metrics despite not being perceptually significant, a discrepancy better addressed by perceptual MOS-based evaluations such as DNSMOS.

From the reported MCD reconstruction metric for synthesizers, we can observe a large improvement by using the DDSP Vocoder. Note that this metric, although intrusive, performs several processing steps to ensure alignment and normalization between the generated and reference signals, as described in subsection~\ref{sec:fig-of-merit}.

Lastly, we observe that the CTN alone, without the DDSP vocoder, can benefit from complex-STFT inputs, in particular for small-size models. However, the same trend is not observed when comparing the results obtained with the DDSP vocoder. We hypothesize that, since the DDSP vocoder itself is a zero-phase filter model, carrying over this information does not improve model performance and slightly increases computational cost, thus not being the best design choice in our experiments.

\begin{table}[!htpb]
    \centering
    \caption{Mean and standard deviation (across samples) results of the proposed DDSP vocoder for speech enhancement on a low-resource and causal CRN model.}
    \label{tab:causal_experiments}
    \resizebox{\linewidth}{!}{%
        \begin{tabular}{lcccc}
            \toprule
            Metrics & CRN & (\textbf{Ours}) CRN + DDSP \\
            \midrule
            \multicolumn{3}{c}{Small version ($\approx$300K parameters)} \\
            \midrule
            PESQ ($\uparrow$) & 1.83 \textcolor{gray}{$\pm$ 0.29} & \textbf{1.88} \textcolor{gray}{$\pm$ 0.29} \\
            STOI ($\uparrow$) & 71.95 \textcolor{gray}{$\pm$ 8.50} & \textbf{72.96} \textcolor{gray}{$\pm$ 8.42}  \\
            DNSMOS {\scriptsize (OVRL)} ($\uparrow$) & 2.79 \textcolor{gray}{$\pm$ 0.23} & \textbf{3.15} \textcolor{gray}{$\pm$ 0.25}  \\
            MCD ($\downarrow$) & 6.91 \textcolor{gray}{$\pm$ 1.27} & \textbf{5.92} \textcolor{gray}{$\pm$ 0.95}  \\
            SSIM ($\uparrow$) & 0.81 \textcolor{gray}{$\pm$ 0.05} & \textbf{0.83} \textcolor{gray}{$\pm$ 0.04}\\

            \midrule
            \multicolumn{3}{c}{Large version ($\approx$600K parameters)} \\
            \midrule
            PESQ ($\uparrow$) & \textbf{1.85} \textcolor{gray}{$\pm$ 0.30} & 1.81 \textcolor{gray}{$\pm$ 0.28} \\
            STOI ($\uparrow$) & \textbf{72.75} \textcolor{gray}{$\pm$ 8.59} & 72.36 \textcolor{gray}{$\pm$ 8.26} \\
            DNSMOS {\scriptsize (OVRL)} ($\uparrow$) & 2.77 \textcolor{gray}{$\pm$ 0.22} & \textbf{3.12} \textcolor{gray}{$\pm$ 0.23} \\
            MCD ($\downarrow$) & 6.86 \textcolor{gray}{$\pm$ 1.28} & \textbf{6.09} \textcolor{gray}{$\pm$ 1.03}\\
            SSIM ($\uparrow$) & 0.81 \textcolor{gray}{$\pm$ 0.05} & \textbf{0.82} \textcolor{gray}{$\pm$ 0.04}\\

            \midrule[\heavyrulewidth]
            \bottomrule
        \end{tabular}
    }
\end{table}

\subsection{Real-time use-case with the causal CRN}

Next, we evaluate the CRN model in a causal setting, emulating a real-time scenario with an algorithmic latency of \SI{8}{ms}, as presented in Table~\ref{tab:causal_experiments}. Our results confirm the feasibility of integrating the DDSP vocoder in a low-latency causal pipeline, with consistent improvements across both intrusive and non-intrusive metrics. However, the observed gains are smaller than in the non-causal configuration. This is expected, as our generative approach relies on sequential prediction of acoustic parameters (e.g., fundamental frequency), which can be more accurately estimated when future context is available, highlighting a key limitation of causal inference in modeling temporally complex speech patterns, such as harmonic structures, that are fundamental for acoustic parameter estimation in vocoder-based methods.

Furthermore, the results indicate that scaling the U-Net-based CRN architecture does not yield significant performance gains. Both the small and large variants, when combined with the DDSP vocoder, achieve comparable scores across all evaluation metrics. Similarly, without the vocoder, increasing the model size yields no significant improvement, suggesting that effectively enhancing the CRN’s performance may require a much larger model. However, such an increase would conflict with the resource constraints of wearable devices. 

\subsection{Scaling-up the CTN model}

Lastly, returning to the CTN model, we investigate the performance gains of scaling up the model's number of parameters. To this end, we slightly change the model sizes for the dual-path Conformer network, in order to have models with roughly 300K, 600K, 1500K, and 2500K parameters.

Figure~\ref{fig:ablation_size_experiments} presents the model size (in number of parameters) on the X-axis and the evaluation metrics on the Y-axis. Moreover, the models augmented with the DDSP vocoder show a strong performance even at small scales, particularly for DNSMOS and MCD. Additionally, when comparing the smallest and largest models, the DDSP-based approach yields a greater relative improvement than the baseline, indicating its efficiency in scaling performance with model size.

\begin{figure}[H]
        \centering
        \includegraphics[width=\linewidth]{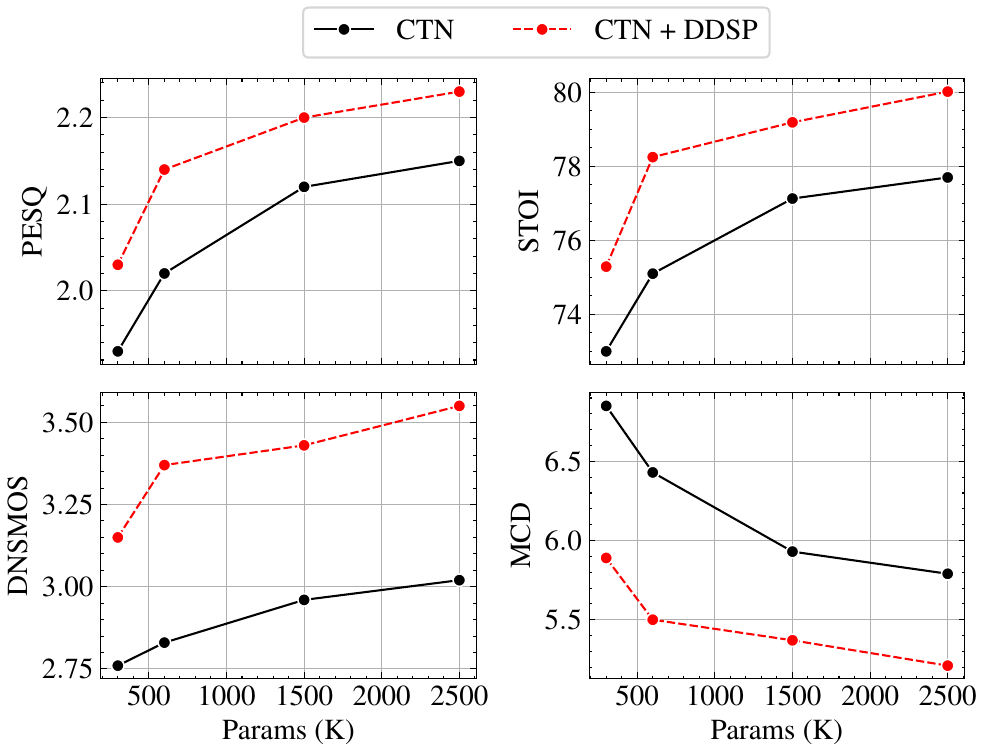}
        \caption{CTN (non-causal) performance across model scales. Results highlight the benefits of the proposed DDSP vocoder and show that increasing model size, when computational resources allow, improves quality and intelligibility.}
        \label{fig:ablation_size_experiments}
\end{figure}

\section{Conclusion}

In this work, we proposed a new approach to improve speech enhancement performance by integrating a differentiable digital signal processing (DDSP) vocoder for synthesizing enhanced utterances: a source-filter-based model that introduces an inductive bias grounded in speech production into the enhancement pipeline. Our approach consists of a compact backbone network that estimates clean acoustic parameters from noisy inputs (spectral envelope, fundamental frequency, and periodicity), followed by a zero-phase DDSP vocoder that reconstructs the waveform. The system is trained end-to-end using a combination of feature- and waveform-level losses within an adversarial learning framework, where the enhancement model acts as the generator. We evaluated our method on a subset of the DNS 2020 dataset using low-resource model configurations, and demonstrated consistent improvements in both intrusive and non-intrusive metrics. In addition, we show the feasibility of the approach under real-time, low-latency constraints in a causal setting with a \SI{8}{ms} algorithm latency. Overall, our findings support the integration of DDSP vocoders into resource-efficient enhancement systems and highlight their potential for edge deployment.

\bibliographystyle{IEEEtran}
\bibliography{references}

\end{document}